# Cloud BI: Future of Business Intelligence in the Cloud

Hussain Al-Aqrabi*, Lu Liu*, Richard Hill, Nick Antonopoulos
*Distributed and Intelligent Systems Research Group, School of Computing and Mathematics, University of Derby, Derbyshire, United Kingdom*

ABSTRACT

Cloud computing is gradually gaining popularity among businesses due to its distinct advantages over self-hosted IT infrastructures. Business Intelligence (BI) is a highly resource intensive system requiring large-scale parallel processing and significant storage capacities to host data warehouses. In self-hosted environments it was feared that BI will eventually face a resource crunch situation because it will not be feasible for companies to keep adding resources to host a never ending expansion of data warehouses and the online analytical processing (OLAP) demands on the underlying networking. Cloud computing has instigated a new hope for future prospects of BI. However, how will BI be implemented on cloud and how will the traffic and demand profile look like? This research attempts to answer these key questions in regards to taking BI to the cloud. The cloud hosting of BI has been demonstrated with the help of a simulation on OPNET comprising a cloud model with multiple OLAP application servers applying parallel query loads on an array of servers hosting relational databases. The simulation results have reflected that true and extensible parallel processing of database servers on the cloud can efficiently process OLAP application demands on cloud computing. Hence, the BI designer needs to plan for a highly partitioned database running on massively parallel database servers in which, each server hosts at least one partition of the underlying database serving the OLAP demands.

---------------------------

## 1. Introduction

Cloud computing has become one of the revolutionary technologies over recent years. Cloud computing is conceptualized in three forms – software-as-a-service (SaaS), platform-as-a-service (PaaS) and infrastructure-as-a-service (IaaS). The SaaS providers interface with the end users by virtue of provisioning of business application services similar to the ones that have been traditionally self-hosted by the corporate houses [1]. Cloud computing paradigm has emerged to bring large-scale computing, storage resources, and data service resources together to build a VCE (virtual computing environment) [2]. Cloud computing users can discard the hassles of large-scale investments in hardware and software platforms, in upgrading them regularly and in expensive licenses of application software used to run business processes, related transactions and decision-support systems [3].

The cloud is generally a multi-tenant computing environment; the multi-tenant cloud solutions can optimize resource sharing while providing isolation solution at different levels required to the tenant [4].

------------------

* Corresponding author.
*E-mail addresses*: h.al-aqrabi@derby.ac.uk (H. Al-Aqrabi), l.liu@derby.ac.uk (L. Liu), r.hill@derby.ac.uk (R. Hill), n.antonopoulos@derby.ac.uk (N. Antonopoulos)

This model has ensured better affordability of the best possible application systems thus supporting an increase in efficiency of businesses. [5] Resources are allocated to end users against service requests made by their end terminals, and resources are allocated by a service provisioning engine that verifies the eligibility of users from a separate schema object that holds multi-tenancy data about all cloud users and groups. Once the eligibility is verified, the resources are reserved for the user through session bindings until the computing processes are in progress by the user terminal. The terminal is normally a virtualized client presented through a virtual server farm. However, there can be direct loading of resources as well (example, for data backup). A separate layer monitors the session usage and utilization of resources such that the billing related information can be generated [6]. NIST (National Institute of Standards and Technology) is in the process of developing standard protocols for user connectivity to the cloud through virtualization interface, terminal emulation interface, thin client interface and Internet browser interface. As of now, there is no standard protocol for users' connectivity to cloud hosted resources [7].

Business intelligence (BI) has been historically one of the most resource intensive applications. It comprises a number of data warehouses created by fetching decision-support data from organization wide databases. The data warehouses are updated at frequent intervals through appropriate queries executed on the business processing and transactional databases. online analytical processing (OLAP) is the user-end interface of BI that is designed to present multi-dimensional graphical reports to the end users. OLAP employs data cubes formed as a result of multidimensional queries run on an array of data warehouses. Furthermore, an OLAP application fetches data from the data warehouses, organizes them in highly complex multidimensional data cubes, and presents to the users through user defined and configured GUI dashboards [8]. BI and OLAP framework has a high business utility, because it helps in locating and eliminating or solving business process deficiencies, inefficient process steps and waste process steps. A BI and OLAP framework is expected to provide timely, accurate, organized and integrated information to business decision makers [8][9].

Despite of excellent business utility of BI and OLAP framework, many business owners were compelled to look for its alternative because of uncontrolled increase in computing and storage resource requirements in self hosted environments. At some stage, the cost of maintaining and upgrading the BI and OLAP framework becomes unjustified for a business [10]. However, the unique selling points of cloud computing offer exactly what businesses need to successfully run BI and OLAP frameworks-unlimited resources, resource elasticity (resources on demand), moderate usage costs, high uptime and availability, high security, no hassles of upgrading and maintaining loads of servers and databases, and so on [1][5]. Hence, it is hereby argued that cloud computing has the potential to offer a new lease of life on BI and OLAP framework. Moreover, it is also argued that cloud computing can extend the power of BI and OLAP to small- and medium-scale businesses, which could not have afforded the framework in self-hosted IT infrastructures. However, it is important to establish a framework for implementing BI and OLAP on cloud computing platform.

The rest of the paper is organized as follows. Section 2 is a literature review on how BI and OLAP framework can be implemented on the clouds and presents key benefits of cloud computing for BI. Section 3 shows an approach for taking BI to the cloud as well as key challenges in hosting BI on cloud. Section 4 describes and explains in detail how BI and OLAP framework can be modeled on a cloud and how it should behave in order to extend maximum utility to the businesses by virtue of an OPNET based simulation experiment. Section 5 presents the summary of the research results and analysis. Finally, the conclusions of this work with future directions are discussed in Section 6.

## 2. Literature reviews

*2.1. A review of Business Intelligence and OLAP and their porting on cloud computing*

Business intelligence (BI) is employed for monitoring the performance of business processes through accurate presentation and analysis of multidimensional data taken from distributed transaction processing systems across the enterprise [23]. The analysts using OLAP-enabled dashboards and reporting systems prefer to map financial data with performance data (of people and processes) to identify inefficiencies and reduce them through strategic restructuring of the business processes and workflows [23]. As per [24], a BI system is made of seven layers: IT and related infrastructure, data acquisition, data integration, data storage, data organising, data analytics and data presentation.

The OLAP cubes form the data storage (partially, in the form of views) and data organizing layers of a BI system. It sits over the data warehouse tables in the form of multidimensional views. A cube structure is made of a number of cross-referenced columns having data fetched from different processes and financial data tables over a period. The periods in a cube are shorter than the ones in the data warehouse that is tasked to store process related and financial data of longer periods (typically, five years or more). The OLAP queries are heavy duty search commands comprising data from multiple views at a time. The presentation layer makes the data sets visible in the form of multidimensional graphical screens [25,26].

A number of OLAP query types are used to generate the views and present them in the dashboards/presentation screens. Some of the popular OLAP queries include: slice and dice, pivoting queries, merge/split queries, rolling-up queries, and drilling down queries [26]. The OLAP cube can be visualized as a stack of two-dimensional matrix planes, whereby each matrix plane represents a relationship between two different dimensions [26]. Figure 1 is a presentation of a multidimensional OLAP cube comprising two dimensional matrix planes.

It should be noted that these matrices are not independent of each other. All the attributes in these planes are nested with each other, and are linked with a primary key that controls all the relationships across multiple composite elements many-to-many relationships. For example, a product code can be viewed as a primary key that controls the relationships among all the attributes related to sales in a business [27].

Formation of OLAP cubes in an enterprise is not an easy task. A well planned architecture needs to be in place for content integration, modelling/mapping and presentation. The typical architecture should comprise of: components and connectivity for federated access to all the and DSS (decision support systems) across the enterprise, a data dictionary designed as per the OLAP cube formulations, a metadata mapping as per the data dictionary, a real data repository (warehousing), a virtual repository (data views serving as building blocks of the OLAP cubes) and the advanced data presentation services (dashboards and customised BI reporting interfaces). [28] Accurate data modelling, metadata mapping and content integration helps in accurate formation of multidimensional cubes and finally resulting in accurate presentation of business issues [29]. The data presentation should be done in such a way that the human-computer interfacing is as friendly as possible (sorting, charting, color coding and multidimensional interactive features) [30].

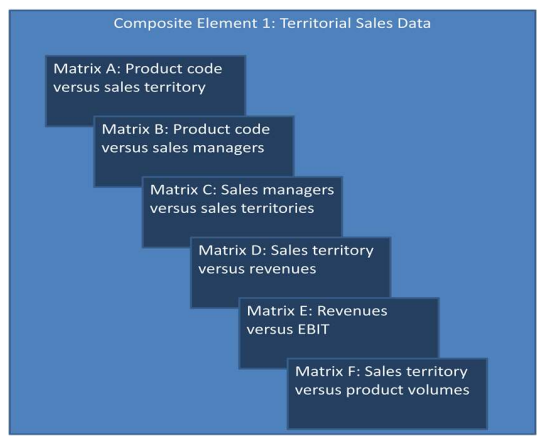

**Fig. 1.** A matrix view of a multidimensional OLAP cube

The Document Type Data (DTD) definition language for defining an XML schema is the primary enabler for taking BI to the clouds. The multidimensional data views in the OLAP cubes can be formed by including DTD parsed XML data files. DTD parsed XML files in a OLAP cube results in the right structures to make use of web services oriented architecture of a cloud. The dashboard/BI reporting interfaces and data analytics layers are built in applications that can be hosted as SaaS mode of cloud hosting. The data warehouses and OLAP cubes can be formed employing multidimensional and hierarchical XML data files. The one-to-many DTD structures can be used to create the data warehouses and the many-to-many DTD structures can be used to create the OLAP cubes. The cubes and warehouses can be hosted on the PaaS mode of cloud hosting. The underlying servers, databases, storage and networking infrastructure components can be hosted on the IaaS mode of cloud hosting [22].

*2.2. BI and OLAP*

BI and OLAP framework comprises a highly complex multi-layer structure. Following are the key components of BI and OLAP framework [8]:

- A user interface layer comprising a large library of dashboards for graphical reporting.
- A layer for data analytics comprising what-if scenarios, reports, stored queries and data models.
- A layer for storing the OLAP cubes formed by multi-dimensional data extraction from the data layer (the data warehouses).
- A data integration layer for identification, cleaning, organizing and grouping of data extracted from the data warehouses before the cubes are formed.
- A data layer comprising of the data warehouses.
- A layer for acquiring data from the business processing, decision support and transactional databases used by various functions of the organization.
- The layer comprising the IT infrastructure components and related resources (data processing, storage and networking).

The key feature of a BI and OLAP framework is the OLAP cube, which is a multidimensional view formed in the structure of a matrix. The OLAP cube is a complex data view formed by running simultaneous queries on the tables of the underlying data warehouses that fetch at least three times more data compared with an ordinary database query. Each cube comprises a stack of multiple two dimensional reports (an ordinary planar graph showing a relationship between two variables). In typical OLAP applications, the queries fetch typically 10 to 12 times more data than an ordinary database query [11]. An OLAP application may comprise multiple OLAP cubes stored in the form of a complex hierarchy of matrices having data organized in the form of cross-tabulations. The cubes are normally stored in separate data marts or within predefined tables in the data warehouses. [12] The common OLAP functions employed for formation of such cubes with a hierarchy of cross-tabulated data are drill-down, merge/split, roll-up, slice-and-dice and pivoting. Each matrix plane is identified by its own classification comprising different data mappings. The planes form a nest-like structure due to interrelationships. The resulting relationship looks like a tree with the roots comprising the primary variables and the branches comprising the secondary variables. For example, a product code is a primary variable and revenues generated in a sales location is a secondary variable. The dashboard operator can modify or change the primary and secondary variables, which directs the query to fetch a different set of data to form different cross-tabulations in the next querying cycle on the underlying data warehouses. Hence, the OLAP cubes are flexible and can be changed dynamically as per the business needs [13]. The following figure shows the BI and OLAP framework:

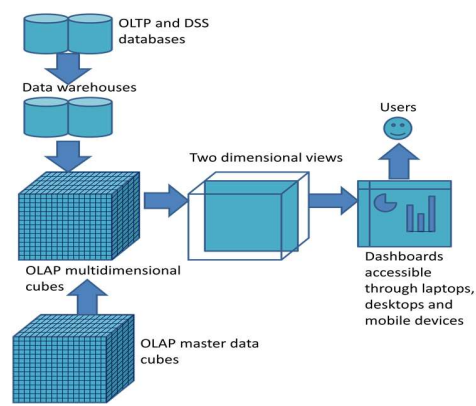

**Fig. 2.** The BI and OLAP framework

The figure shows two forms of cubes–the OLAP multidimensional data cubes and the OLAP master data cubes. The master data cubes control the relationship formation between the two-dimensional data planes within the multi-dimensional data cubes. Business users are offered a large range of variables that they can combine to form different views of two-dimensional reports needed in the dashboards. The data is pulled from OLTP (online transaction processing) and DSS databases into the data warehouse tables periodically, which in turn helps in periodic automatic updating of the data in the data cubes and finally in the dashboards. Hence, the business users can closely monitor business performance by virtue of updating dashboards continuously. Appropriate color coding of reference points/thresholds helps in generating alerts and alarms that helps the business strategic decision-makers to take appropriate steps [14,5].

*2.3. Benefits of Cloud BI*

Nowadays, Cloud BI solutions are gradually gaining popularity among businesses, as many businesses are realizing the benefits of data analytics. Businesses need quality insights driven by accurate data more than ever. The SaaS providers are serving as the primary interfacing to the business users community [22]. Cloud BI is the concept of delivering BI capabilities as a service. The following are key benefits of cloud computing for business intelligence.

**Cost efficiency**

In the cloud, companies do not need to budget for large, up-front purchases of software packages or carry out time-consuming updates on local servers to put the BI infrastructure up and running. They will treat it as a service, paying only for the computing resources they need and avoids costly asset acquisition and maintenance reducing the entry threshold barrier.

**Flexibility and Scalability**

Cloud BI solutions allows for greater flexibility to be altered quickly to give technical users access to new data sources, experimenting with analytical models and easy to keep fiscal control over IT projects and have the flexibility to scale up or down usage as needs change. Moreover, In the cloud, resources can automatically and rapidly scale in and scale out, and it can support large numbers of simultaneous users. This means that customers can easily increase their software usage without delay or the cost of having to deploy and install additional hardware and software.

**Reliability**

Reliability improves through the use of multiple redundant sites, which can provide reliability and secure locations for data storage and the resources can be spread across a large number of users , which makes cloud computing suitable for disaster recovery and business continuity.

**Enhanced data sharing capabilities**

Cloud applications allow data access to be shared remotely and enable easy cross-location data sharing capabilities as they are deployed via the internet and outside a company's firewall.

**No capital expenditure**

Low TCO (total cost of ownership) is a key benefit of the cloud model. With the cloud, companies pay for a service they actually use. With this policy, cloud computing allows companies to better control the CAPEX (Capital Expenditure) and the OPEX (Operations Expenditure) associated with non-core activities. Hence, the benefits of BI can be rolled out faster to more users within the organization.

## 3. Taking BI to the cloud

BI in the cloud is a game changing phase of IT, as it makes BI finally affordable and accessible as compared to traditional BI. On the cloud, the matrices in the OLAP cubes can be formed using the web data warehousing concept making use of XML data files using DTD (document type definition) described XML programming language. The data structures in the cubes are formed using the DTD parsed XML files [14]. The DTD format helps an XML file to exhibit relational properties of a conventional database. This is what enables the OLAP cubes stored on the cloud making use of XML data files following DTD structures (called web cubes). This also helps the BI system to make use of web services components thus ensuring better performance on the cloud [17][18]. The entire OLAP framework comprising the dashboards and the data analytics layer can be hosted as SaaS. The BI and OLAP framework software platforms available for cloud hosting are SAP, IBM Cognos and Web-Sphere Dashboards dashboard, Oracle business objects and Salesforce.com. The integration of data warehouses (XML based) and OLTP/DSS databases can be hosted on PaaS. The underlying servers and databases can be hosted on IaaS mode of cloud hosting. For optimum performance on the cloud, the servers and database arrays should be implemented as a massively parallel system capable of processing large scale parallel queries [19].

The databases on the cloud needs to be implemented in the form of a massively parallel system to support high demand elasticity of BI and OLAP framework. A centralized schema object may be designed to hold the details and privileges of all tenants on the cloud. Each schema object holding the data files may be massively partitioned such that each partition can be held by a separate server on a large scale server array. The IaaS provider should be capable of rapid expansion of the server array making use of virtualized array expansion. In this way, it may be possible to serve one partition through more than one servers that can enhance the performance of BI. The IaaS provider should keep a close watch on both load distribution and response time patterns and make effective network changes to ensure that the network load is also distributed evenly [21]. The OLAP application hosted on the cloud may not be web services compatible. To make an OLAP application compatible to web services architecture, the SaaS provider may allow the creation of an intermediate layer to host a dependency graph that helps in dropping the attributes not needed in the finalized XML data cube [20,21].

Hence, the following are key challenges in hosting BI on cloud:

- Compliance of the BI application with web services architectural standards (and the standards defined by the SaaS or PaaS provider, like Google Apps standards)
- Deployment of massively parallel data-warehousing system with evenly distributed query load and even patterns of response times from all database servers. The IaaS provider should effectively use the virtualized server array management and expansion to meet the resources on demand.
- The network architecture should be designed in such a way that the query load can be evenly distributed among the servers in an array. This will ensure even query processing response times by the servers in an array. If the server array employs storage area networking for storing the XML data files and the OLAP cubes, the data fetching from various storage devices should again be evenly distributed by virtue of appropriate network connections.

In the next section, an OPNET model of a small scale BI and OLAP framework on the cloud has been created. The network has been designed in such a way that the load can be evenly distributed to all the relational database management systems (RDBMS) servers. In addition, the application demands have been created in such a way that all RDBMS servers are evenly involved in receiving and processing the OLAP query load. The BI on the cloud model is described in the next section and the simulation results are described in the subsequent section.

## 4. BI on the cloud model

This section provides a brief description of the main interface of the OPNET model. The model comprises two large domains–the BI on the cloud domain and the Extranet domain comprising six corporates having 500 OLAP users in each.

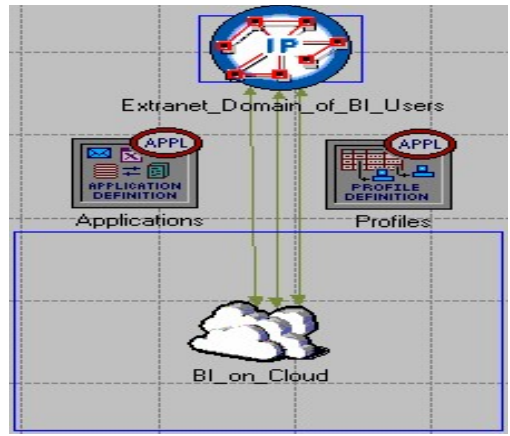

**Fig. 3.** The architecture of the model

The BI on the cloud domain is expanded in Figure 4 below. The cloud has been formed using four numbers of Cisco 7609 high end routing switches connecting in such a way that the load can be evenly distributed. The cloud switch 4 is dedicated to route all inbound traffic to the servers and send their responses back to the clients. The cloud switches 1 and 3 are serving four RDBMS servers each and the cloud switch 2 is serving all the OLAP application servers.

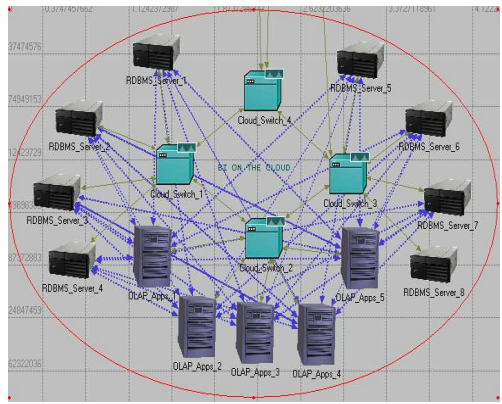

**Fig. 4.** The BI on the cloud architecture

The blue dotted lines indicate the traffic flow distribution from the OLAP application servers to the RDBMS servers. As shown in the figure, the load from the OLAP application servers are evenly distributed among all RDBMS servers. The client load is routed to the OLAP application servers using destination preference settings on the client objects configured in the extranet domain, as shown in Figure 5 below.

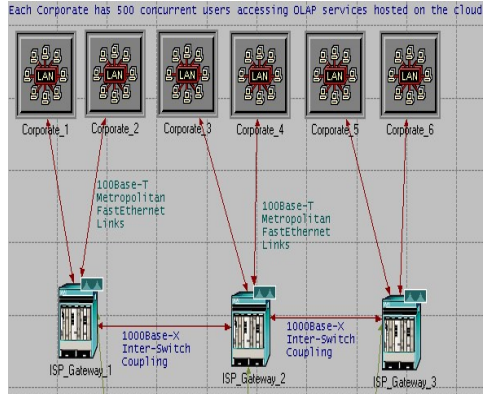

**Fig. 5.** The Extranet domain comprising six corporates having 500 OLAP users in each corporate

The extranet comprises of three ISP gateway switches serving six corporate LAN segments having 500 users each. Each LAN objects have the four OLAP servers configured as destination preferences for the OLAP application profile. In this way, the OLAP requests from the clients are routed to the four OLAP servers and the RDBMS requests are routed from the four OLAP servers to the eight RDBMS servers (serving as a small scale server array in this model).

The RDBMS queries are configured using the attributes shown in Table 1 below. The default configurations of heavy database load of OPNET has been chosen and then increased by 10 times in the table below. This is based on the literature review that OLAP query load on databases is at least 10 times heavier than the normal query load. Moreover, the inter-arrival time of query has been set at one second, and the type of service has been set at "excellent service". Finally, the transaction mix of queries versus total transactions has been set at 100%. This is because the BI and OLAP framework does not has any data entry load because the framework is used for strategic decision support.

**Table 1**
The Database Query Settings To Emulate OLAP Query Load On The Databases

| Attribute | Value |
| --- | --- |
| Transaction Mix (Queries/Total Transactions) | 100% |
| Transaction Interarrival Time | constant (1) |
| Transaction Size (bytes) | constant (10240) |
| Symbolic Server Name | RDBMS Server |
| Type of Service | Excellent Effort (3) |
| RSVP Parameters | None |
| Back-End Custom Application | Not Used |

The OLAP application has been configured as a heavy browsing HTTP application having varying 5120 bytes to 10240 bytes of object downloads per second (continuously updating dashboards), 7 to 10 objects per interface (dashboards, its description screens, legends, text boxes, and so on), one second object refresh time (because the transaction inter-arrival time on the databases is one second) and 10-second page refresh time (ensuring that the OLAP screen refreshes after every 10 cube refreshes such that the user gets noticeable data changes at every screen refresh).

**Table 2**
The OLAP Application Profiling

| Attribute | Value |
| --- | --- |
| name | Profiles |
| model | Profile Config |
| Profile Configuration | (...) |
|   rows | 1 |
|   row 0 | |
|     Profile Name | OLAP_Requests |
|     Applications | (...) |
|       rows | 2 |
|       row 0 | |
|         Name | OLAP |
|         Start Time Offset (seconds) | uniform (5,10) |
|         Duration (seconds) | End of Profile |
|         Repeatability | (...) |
|           Inter-repetition Time (sec... | exponential (300) |
|           Number of Repetitions | Unlimited |
|           Repetition Pattern | Concurrent |
|       row 1 | |
|         Name | RDBMS_Services |
|         Start Time Offset (seconds) | uniform (5,10) |
|         Duration (seconds) | End of Profile |
|         Repeatability | (...) |
|           Inter-repetition Time (sec... | exponential (300) |
|           Number of Repetitions | Unlimited |
|           Repetition Pattern | Concurrent |
|   Operation Mode | Simultaneous |
|   Start Time (seconds) | uniform (50, 55) |
|   Duration (seconds) | End of Simulation |

The application profiling of OLAP application (OLAP requests) and the RDBMS services. Both the profiles trigger concurrently with an offset of 5 to 10 seconds after the start time. The start time has been configured at 50 to 55 seconds to ensure that all routing updates are successfully completed on the network before the application services are triggered.

**5. Research Results and Analysis**

In this research, the results shown here are the ones captured from a simulation of 50 million events which is the maximum possible in OPNET academic edition. The query load is not exactly the same on the RDBMS servers but the pattern indicates almost even distribution of query load. This is evident from the "database query requests per second" statistics collected from the eight RDBMS servers stacked one above another as shown in Figure 5 below. The query requests experienced by each server are plotted in the form of "number of queries per second" on the Y-axis with respect to simulation time on the X-axis. In Figure 6, this statistic is reported for RDBMS server 1 through server 8 on the cloud.

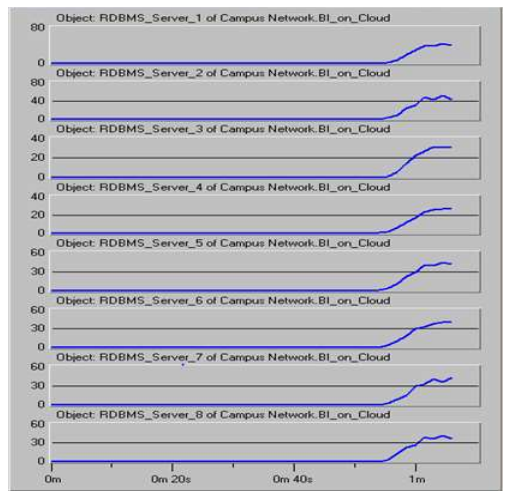

**Fig. 6.** Query load on the RDBMS servers

The query load is slightly above or below 40 requests per second on all the RDBMS servers. This reveals that the load distribution through appropriate network configuration and application demand profiling (traffic flow configurations indicated by blue dotted lines in Figure 4). These configurations have caused near even distribution of query load from the four OLAP servers on the RDBMS servers. Moreover, the query task processing times on the database servers are also nearly even as shown in Figure 7 below. This has been possible because the same hardware make, model and configurations has been chosen for all the eight RDBMS servers. The query response time by each server is plotted in the form of "number of queries processed per second" on the Y-axis with respect to simulation time on the X-axis. In the Figure 7, this statistic is reported for RDBMS server 1 through server 8 on the cloud.

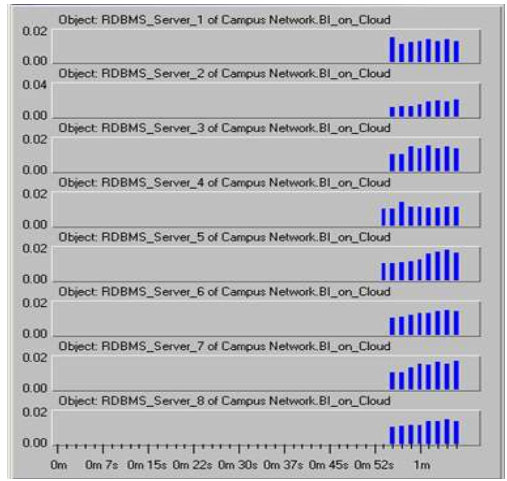

**Fig. 7.** Query task processing time by the RDBMS servers.

These results are a good demonstration of how a massively parallel RDBMS system can be deployed to form a BI and OLAP framework and how the framework should perform in the cloud environment. This is in line with the requirements stated by researchers as reviewed in the literature study. However, there are a few key points that should be kept in mind about this model as listed below:

First, the model has only eight servers in the RDBMS array serving only four numbers of OLAP application servers.
Second, the load distribution has been managed evenly through application demand flow modeling which is an excellent feature of OPNET and works very well.
Third, the servers chosen in this model are of the same make and model having identical hardware configuration.
Fourth, the load has been modeled as constant after an exponential increase at the start. The simulation of the load carried out in this model has lasted only for 50 million events and with no load variations.
Finally, this model comprises only 3000 OLAP users connecting concurrently. A real BI environment on cloud computing will have tens of thousands of end users applying concurrent BI load on the servers.

These are ideal scenarios that will not be possible on the real cloud. However, these settings in OPNET has evolved into challenges that will be faced in moving BI to the cloud as per the requirements stated by the researchers. A cloud will have hundreds of servers in the arrays; hence, even distribution of network load will be a very challenging task.

The architects will have to watch for bottlenecks on the inter-switch connections, even if they are deployed using the fastest possible ATM connections or the 10G gigabit Ethernet. The load distribution will have to be managed by advanced provisioning engines and routers, which will not be as easy as configuring application demand flow patterns in OPNET as indicated by blue dotted lines in Figure 4. These provisioning engines and routers need to be optimized to ensure that the user load is evenly distributed among the servers in the array and spilled over to additional arrays if there is an overloading scenario. The partitioning of databases in the data warehouses should be achieved in such a way that the arrays can be quickly expanded and new servers can start contributing resources in serving the partitions without carrying out any structural changes of the databases. At some stage, it should be made possible to deploy both the data warehouses and OLAP cubes employing XML data files, thus completely eliminating the need for traditional RDBMS software systems in the BI and OLAP framework.
On the hardware side, it may not be possible for the IaaS provider to implement a cloud with identical hardware make, model and configurations. Hence, the query processing response time of each server will be different on the cloud due to differences in hardware configurations. Hence, a mere even distribution of load to the servers by the service provisioning engine and the router will not serve the purpose. There should be some intelligence to route the load based on the knowledge of query processing response times of the servers. The servers with slower response times should get lesser load compared with the servers with faster response times to eliminate wait states at the receiving end. The capabilities of RDBMS partitioning, RDBMS load balancing, web provisioning application services, services routing engines and query performance optimizing should be exploited effectively by the BI architects. This is to ensure that the massively parallel processing system of database server arrays works perfectly to effectively utilize the processing power of the servers and synchronize the query processing times to reduce or eliminate wait states at the application servers' end.

The above discussion presents one more challenge in taking BI to the clouds. The SaaS, PaaS and IaaS providers may be different companies. Hence, to ensure the above requirements of BI hosting on clouds, these providers need to carry out excellent coordination of architectural detailing for designing and deploying the services to enable the various layers of BI and OLAP framework. BI cannot be implemented in an ad-hoc way by the providers otherwise it will suffer from the same level of bottlenecks and resource crunch as it has been suffering in self-hosted environments. The providers need to carry out effective planning of every detail and implement the infrastructure components, platform components and application components to achieve a true massively parallel processing system with highly elastic capacity enhancement framework using all available technologies efficiently.

## 6. Conclusions and Future Directions

Cloud is a big part of future Business Intelligence (BI) and offers several advantages in terms of cost efficiency, flexibility and scalability of implementation, reliability, and enhanced data sharing capabilities. Cloud has the potential to offer a new lease of life to BI and OLAP framework. Cloud computing comprises three ways of provisioning services – software-as-a-service (SaaS), platform-as-a-service (PaaS) and infrastructure-as-a-service (IaaS). These services may be provided by same or different providers depending upon the business arrangements. However, the SaaS provider needs the settings on the PaaS and IaaS clouds to be defined as per the application services provisioned through the web services architecture components. Clouds comprise the service provisioning and routing engines that can effectively sense the loading pattern on the underlying resources.

BI and OLAP framework is highly resource intensive. it has a multilayer architecture comprising multidimensional OLAP cubes with multiplexed matrices representing relationships between various business variables. The cubes are formed by sending OLAP queries to the data warehouses stored in the RDBMS servers. The size of an OLAP query is typically 10 to 12 times larger than an ordinary database query. Hence, if BI and OLAP framework is taken to the cloud for serving hundreds and thousands of end users, it is essential that the cloud providers implement massively parallel processing RDBMS systems with even distribution of query load and query response times for the OLAP application servers. In this research, a BI and OLAP framework has been modeled using OPNET and the requirements of a massively parallel RDBMS server array has been modeled using the OPNET features. The results have reflected the ideal scenario for taking BI to the cloud. However, the real clouds will not have ideal configurations as made in this OPNET model. Hence, the real challenges on the cloud needs to be identified and addressed to ensure that the results can be brought closer to ideal scenario as far as possible.

The details of challenges in implementing a massively parallel processing RDBMS server system to take BI to the cloud has been discussed. Many settings that are possible in OPNET simulation environment may require significant architectural innovations to achieve what has been described in this paper to successfully take BI to the clouds.

In the future, researchers may like to study modern technologies pertaining to service provisioning, service routing, schema partitioning, load balancing, and so on to implement an enterprise level RDBMS system to achieve a massively parallel processing RDBMS server system for taking BI to the clouds. In this context, there is a significant opportunity to carry out multiple experimental studies to evolve the practical configuration solutions useful for the cloud service providers targeting to host BI and OLAP framework on the cloud.

Cloud computing also offers significant computing power and capacity. Hence, BI is expected to enter many complex domains (business and non-business related) which were impossible for it in a self-hosted environment. Applications like context-aware, location-aware automation, massive scale semantics, advanced science and technology databases, real time disaster and crisis management, city management, global finance and economy reporting, and global monitoring of industries and sectors are few such areas where BI or BI like systems possess tremendous potential on Cloud computing. The size, scale, dynamism, and scope of data marts and data warehouses on Clouds may exceed even the Petabytes scale (the emerging challenge of Big Data). Such data systems cannot be managed using traditional systems and tools. The security challenges at such massive scales will be different and much more complex. Hence, this research has significant openings for future contributions. This concept is a kind of a beginning of setting a stage for studies on such future challenges.